\documentclass[hyper,12pt,letterpaper]{JHEP3}
\usepackage{graphicx}

\title{Spectrum of Dyons and Black Holes in CHL orbifolds
using Borcherds Lift}

\preprint{\hepth{0603066}\\TIFR-TH-06-07}

\author{Atish Dabholkar and Suresh Nampuri \\

\centerline{Department of Theoretical Physics, Tata Institute of
Fundamental Research,}

\centerline{Homi Bhabha Road, Mumbai 400005, India}}

\abstract{The degeneracies  of supersymmetric quarter BPS dyons in
four dimensions and of spinning black  holes in five dimensions in a
CHL compactification are computed exactly using Borcherds lift. The
Hodge anomaly in the construction has a physical interpretation as
the contribution of a single M-theory Kaluza-Klein 6-brane in the
4d-5d lift. Using factorization, it is shown that the resulting
formula has a natural interpretation as a two-loop partition
function of left-moving heterotic string, consistent with the
heuristic picture of dyons in the M-theory lift of string webs.}


\renewcommand{\Im}{\mbox{Im}}

\def\IC{\relax\hbox{$\inbar\kern-.3em{\rm C}$}}

\def\IC{{\bf C}}

\def\bea{\begin{eqnarray}}
\def\eea{\end{eqnarray}}
\def\be{\begin{equation}}
\def\ee{\end{equation}}
\def\ba{\begin{align}}
\def\ea{\end{align}}
\def\bse{\begin{subequations}}
\def\ese{\end{subequations}}
\def\1F1{{}_1\!F_1}
\def\2F0{{}_2\!F_0}

\begin{document}

\section{Introduction}

For heterotic string compactified on $\bf T^6$, there exists a
remarkable formula that gives the exact degeneracies of the dyonic
quarter BPS states in the four-dimensional theory
\cite{Dijkgraaf:1996it, LopesCardoso:2004xf, Shih:2005uc,
Gaiotto:2005hc, Shih:2005he, LopesCardoso:2006bg}. A similar formula
has been proposed also for Type-II string compactified on $\bf{T^6}$
\cite{Shih:2005qf, Pioline:2005vi}.

The  spectrum of dyons encapsulates valuable information about the
nonperturbative structure of the theory. Moreover, when the charge
of the dyon is large, it gravitates to form a black hole. The
logarithm of the degeneracies can then be compared with the
Bekenstein-Hawking-Wald entropy. In favorable situations, one can
hope to make exact comparisons between macroscopic and microscopic
degeneracies beyond leading order as was done  for the electric
states \cite{Dabholkar:2004yr, Sen:2005pu, Sen:2005ch,
Dabholkar:2005dt, Dabholkar:2005by} in an appropriate ensemble
\cite{Ooguri:2004zv}.

It is therefore of considerable interest to see if one can obtain a
similar exact formula for counting dyons in more general
compactifications. An interesting class of models where the
computations are  tractable are the CHL orbifolds
\cite{Chaudhuri:1995fk, Chaudhuri:1995bf} of the heterotic string
that result in ${\cal N}=4$ models in four dimensions with gauge
groups of reduced rank. Toroidally compactified heterotic string
results in a gauge group of rank 28. A CHL compactification is
obtained by orbifolding the heterotic string compactified on $\bf
T^4 \times \tilde{S}^1 \times S^1$ by a $\mathbb{Z}_N$ symmetry
generated by $\alpha T_{1/N}$, where $\alpha$ is an order N symmetry
of the internal CFT of the heterotic string compactified on $\bf
T^4$, and $T_{1/N}$ is an order $N$ translation along the circle
$\bf \tilde{S}^1$. The internal symmetry $\alpha$ has a nontrivial
action on the gauge bosons and hence some combinations of bosons are
projected out in the orbifold theory. Because of the order $N$ shift
that accompanies $\alpha$, the twisted sectors are massive and  no
additional gauge bosons arise in the twisted sector. The resulting
theory then has gauge group with a rank smaller than $28$. Using
string-string duality, this heterotic compactification is dual to an
orbifold of Type-II on $\bf K3 \times \tilde{S}^1 \times S^1$ by a
$\mathbb{Z}_N$ symmetry generated by $\tilde\alpha T_{1/N}$, where
$\tilde\alpha$ is an order $N$ symmetry in the internal CFT of the
$\bf K3$.

The S-duality group of a CHL orbifold is a congruence subgroup
$\Gamma_1(N)$ of the $SL(2, \mathbb{Z})$ S-duality symmetry of the
original toroidally compactified heterotic string theory. As a
result, the dyonic degeneracies are expected to display interesting
differences for various orbifolds. Recently a formula  for the exact
dyonic degeneracies for CHL orbifold was proposed
\cite{Jatkar:2005bh} for the cases $N = 2, 3, 5, 7$  generalizing
the work of \cite{Dijkgraaf:1996it}. Let us  summarize this
proposal. For a $\mathbb{Z}_N$ orbifold, the dyonic degeneracies are
encapsulated by a Siegel modular forms $\Phi_k(\Omega)$ of level $N$
and index $k$ as a function of period matrices $\Omega$ of a genus
two Riemann surface\footnote{ Definitions of various quantities
mentioned here in the introduction are given in $\S{\ref{Siegel}}$
and $\S{\ref{Borcherds}}$.}. We denote each case by  the pair $(N,
k)$. The index $k$ is related to the level $N$ by the relation
\begin{equation}\label{level}
    k = {24\over N+1}- 2.
\end{equation}
The original example of toroidally compactified heterotic string is
included in this list as a special case $(1, 10)$. To obtain the
dyonic degeneracy, Jatkar and Sen define
$\tilde{\Phi}_k(\tilde{\Omega})$ related to $\Phi_k(\Omega)$ by an
$SP(2, \mathbb{Z})$ transformations with a Fourier expansion
\begin{equation}\label{tildephi}
    {1\over \tilde\Phi_k(\tilde \Omega)} ={1\over N\, K}
   \sum_{m,n,p\atop m\ge -1,n\ge -1/N}
   e^{2 \pi i(m \tilde\rho
     + n\tilde\sigma + p\tilde \nu)} g(m,n,p)\, ,
\end{equation}
where $K$ is an appropriate constant and
$\tilde \Omega =
\left(
  \begin{array}{cc}
    \tilde \rho & \tilde \nu \\
    \tilde \nu & \tilde \sigma \\
  \end{array}
\right) $. Consider now  a dyonic state with a charge vector $Q=
(Q_e, Q_m)$ which is a doublet of the $SL(2, \mathbb{R})$. Here
$Q_e$ and $Q_m$ are the electric and magnetic charges that transform
as vectors of the T-duality symmetry $O(r-6, 6; \mathbb{Z})$ for a
rank $r$ CHL compactification with $r =2k +4$. The $Q_e^2$, $Q_m^2$,
and $Q_e \cdot Q_m$ be the T-duality invariant
combinations\footnote{The T-duality group of the CHL model is
actually a subgroup of $O(r-6, 6; \mathbb{R})$ symmetry that we have
used here. The degeneracy formula proposed in \cite{Jatkar:2005bh}
therefore is likely to be valid for a restricted class of dyons. In
general, the degeneracy can also depend on more subtle invariants of
the arithmetic subgroup that cannot be written as invariants of the
continuous Lie group as is in the case of electric states.}. The
degeneracy of dyons $d(Q)$ is then given in terms of the Fourier
coefficients by
\begin{equation}\label{degen}
    d(Q) = g(\frac{1}{2} Q_m^2, \frac{1}{2} Q_e^2, Q_e\cdot Q_m).
\end{equation}

The degeneracy $d(Q)$ obtained this way satisfies three nontrivial
physical consistency checks \cite{Jatkar:2005bh}. It is manifestly
invariant under the S-duality group $\Gamma_1(N)$, it agrees with
the Bekenstein-Hawking-Wald entropy of the corresponding black holes
to leading and the first subleading order \cite{Dijkgraaf:1996it,
LopesCardoso:2004xf}, and finally it is integral as expected for an
object that  counts the number of states.

The task of understanding the dyon spectrum is then reduced to
understanding the physics contained in the Siegel modular forms
$\Phi_k(\Omega)$. In the best understood  case $(1, 10)$, the
modular form $\Phi_{10}(\Omega)$ is the well-known Igusa cusp form
which is the unique cusp form of weight $10$.  In  the original
proposal of Dijkgraaf, Verlinde, Verlinde \cite{Dijkgraaf:1996it},
the relation of this modular form to the counting of dyons was
conjectured based on various consistency checks and a heuristic
derivation of the properties of NS5-brane worldvolume theory
\cite{Dijkgraaf:1996it, Dijkgraaf:2002ac}. This conjecture used in
an essential way  the perturbative string computations in
\cite{Kawai:1995hy} of threshold corrections of heterotic string on
$\bf K3 \times T^2$ where the Igusa cusp form naturally appears. In
\cite{Dijkgraaf:1996xw}, the Igusa form made its appearance in an
apparently unrelated context in connection with the elliptic genus
of the symmetric product of $\bf K3$, which counts the bound states
comprising the D1-D5-P black hole in five dimensions
\cite{Strominger:1996sh}. A new perspective on the dyon counting
formula and a definitive connection between the 4d and 5d formulae
was provided in \cite{Shih:2005uc} using the 4d-5d lift of
\cite{Gaiotto:2005gf}. It related the 4d dyonic degeneracies to 5d
degeneracies of the D1-D5-P black holes computed by the elliptic
genus of the symmetric product of $\bf K3$ providing a physical
re-derivation of the dyon counting formula.

Our objective is to obtain a similar physical derivation of the
spectrum of dyons in  CHL compactifications. Towards this end, we
outline in $\S{\ref{Siegel}}$ a general procedure for deriving the
modular forms $\Phi_k$ using a generalization of Borcherds lift
which we call `Multiplicative Lift'. This lift results in a a Siegel
modular form of level $N$ and weight $k$ of $G_0(N)$ in a product
representation, starting with a weak Jacobi form of weight zero and
index one of $\Gamma_0(N)$. The resulting formula suggests a
physical interpretation using the idea of 4d-5d lift proposed in
\cite{Gaiotto:2005gf, Shih:2005uc}. With this interpretation,   the
modular form $\Phi_k$ that counts dyons in four dimensions is
naturally related to a quantity that counts the bound states of
D1-D5-P in five dimensions. In $\S{\ref{Example}}$ we illustrate
this general procedure for the special case $(2, 6)$ and the Siegel
modular form $\Phi_6$. We discuss various aspects of our physical
interpretation using 4d-5d lift in $\S{\ref{Interpret}}$ and obtain
as a byproduct the exact degeneracies of spinning black holes in CHL
compactifications to five dimensions.

One of the mysterious aspects of the counting formula is the
appearance of the genus two modular group. A physical explanation
for this phenomenon was proposed in \cite{Gaiotto:2005hc} by
relating the dyon degeneracies to string webs. Lifting the string
webs to Euclidean M-theory 5-branes wrapping on $\bf K3$ naturally
results in a genus-two Riemann surface using the fact that M5-brane
on $\bf K3$ is the fundamental heterotic string. We offer a similar
interpretation of our results  in $\S{\ref{Web}}$. We show using
factorization that the product formula for $\Phi_6$ has a natural
representation as a chiral, genus-two partition function of the
left-moving heterotic string.

Details of the construction of the  modular forms $\Phi_k$ for the
remaining pairs $(N, k)$ using multiplicative lift   will be
presented in a forthcoming publication along with a  more complete
discussion of the physical interpretation outlined in this note
\cite{Dabholkar:2006}.

\section{Siegel Modular Forms of Level $N$ \label{Siegel}}

Let us recall some relevant facts about Siegel modular forms. Let
$\Omega$ be the period matrix of a genus two Riemann surface. It is
given by a $(2\times 2)$ symmetric matrix with complex entries
\begin{equation}\label{period}
   \Omega = \left(
              \begin{array}{cc}
                \rho & \nu \\
                \nu & \sigma \\
              \end{array}
            \right)
\end{equation}
satisfying
\begin{equation}\label{cond1}
    \Im{(\rho)} > 0,\quad \Im{(\sigma)} > 0,\quad \Im{(\rho)}\, \Im{(\sigma)}
> \Im{(\nu)}^2,
\end{equation}
and parametrizes  the `Siegel upper half plane' in the the space of
$(\rho, \nu, \sigma)$. There is a natural symplectic action on the
period matrix by the group $Sp(2, \mathbb{Z})$ as follows. We  write
an element $g$ of $Sp(2, \textbf{Z})$ as a $(4\times 4)$  matrix in
a block-diagonal form as
\begin{equation}\label{sp}
   \left(
  \begin{array}{cc}
    A & B \\
    C & D \\
  \end{array}
\right),
\end{equation}
where $A, B, C, D$ are all $(2\times 2)$ matrices with integer
entries. They  satisfy
\begin{equation}\label{cond}
   AB^T=BA^T, \qquad  CD^T=DC^T, \qquad AD^T-BC^T=I\, ,
\end{equation}
so that $g^t J g =J$ where $J = \left(
                                  \begin{array}{cc}
                                    0 & -I \\
                                    I & 0 \\
                                  \end{array}
                                \right)$
is the symplectic form. The action of $g$  on the period matrix is
then  given by
\begin{equation}\label{trans}
    \Omega \to (A \Omega + B )(C\Omega + D ) ^{-1}.
\end{equation}
The $Sp(2, \mathbb{Z})$ group is generated by the following three
types of  $(4 \times 4)$ matrices with integer entries
\begin{eqnarray}\label{egroup}
   g_1(a,b,c,d) &\equiv& \pmatrix{ a & 0 & b & 0 \cr
     0 & 1 & 0 & 0\cr c & 0 & d & 0\cr 0 & 0 & 0 & 1}\, ,  \quad ad-bc=1,
   \nonumber \\
   g_2 &\equiv& \pmatrix{0 & 1 & 0 & 0 \cr -1 & 0 & 0 & 0\cr
     0 & 0 & 0 & 1\cr 0 & 0 & -1 & 0}\, , \nonumber \\
   g_3(\lambda, \mu) &\equiv& \pmatrix{ 1 & 0 & 0 & \mu \cr
     \lambda & 1 & \mu & 0\cr 0 & 0 & 1 & -\lambda\cr
     0 & 0 & 0 & 1}\, .
\end{eqnarray}
We are interested in a subgroup by $G_1(N)$ of $Sp(2, \mathbb{Z})$
generated by the matrices in (\ref{egroup}) with the additional
restriction
\begin{equation}\label{G}
   \hbox{$c=0$ mod $N$,
 \quad $a,d=1$
     mod $N$}.
\end{equation}
Note that with the restriction (\ref{G}), the elements $g_1(a, b, c,
d)$ generate the congruence subgroup $\Gamma_1(N)$ of $SL(2,
\mathbb{Z})$ which is the reason for choosing the name $G_1(N)$ for
the subgroup of $Sp( 2, \mathbb{Z})$ in this case. From the
definition of $G_1(N)$ it follows that if
\begin{equation}\label{belong}
    \pmatrix{A & B\cr C & D}\in G_1(N),
\end{equation}
then
\begin{equation}\label{GN}
   \hbox{$C={\bf 0}$
     mod $N$, \quad $\det A = 1$
     mod $N$, \quad $\det D=1$ mod $N$
   }
   \, .
\end{equation}
One can similarly define $G_0(N)$ corresponding to $\Gamma_0(N)$ by
relaxing the condition $a, d =1$ mod $N$ in (\ref{G}).

We are interested in a modular form $\Phi_k(\Omega)$  which
transforms as
\begin{equation}\label{phi}
    \Phi_k [(A \Omega + B )(C\Omega + D ) ^{-1}] =  \{\det{(C\Omega + D )}\}^k
    \Phi_k (\Omega),
\end{equation}
for matrices
$
\left(
  \begin{array}{cc}
    A & B \\
    C & D\\
  \end{array}
\right)
$
belonging to $G_1(N)$. We will actually construct modular forms of
the bigger group $G_0(N)$. Such a modular form is called a Siegel
modular form of level $N$ and weight $k$. From the definition
(\ref{phi}) it is clear that a product of two Siegel modular forms
$\Phi_{k_1}$ and $\Phi_{k_2}$ gives a Siegel modular form $\Phi_{k_1
+ k_2}$. The space of  modular forms is therefore a ring, graded by
the integer $k$. The graded ring of Siegel Modular forms for
$N=1,2,3,4$ is determined in a number of papers in the mathematics
literature \cite{Igusa:ig, Igusa:ig2, Ibukiyama:1991ibu3,
Hayashibu:2005sie, AI:2005ai1}. The special cases of our interest
for the pairs $(N,  k)$ listed in the introduction were constructed
explicitly in \cite{Jatkar:2005bh}.

In the theory of Siegel modular forms, the weak Jacobi forms of
genus one play a fundamental role. A weak Jacobi form
$\phi_{k,m}(\tau, z)$ of $\Gamma_0(N)$ transforms under modular
transformation
$
\left(
  \begin{array}{cc}
    a & b \\
    c & d \\
  \end{array}
\right) \in \Gamma_0(N)
$
as
\begin{equation}\label{modular}
\phi_{k,m}(\frac{a\tau+b}{c\tau+d},\frac{z}{c\tau+d})=(c\tau+d)^k
\exp{[\frac{2\pi i mc z^2}{c\tau+d}]}\,\phi_{k, m}(\tau,z).
\end{equation}
and under lattice shifts as
\begin{equation}\label{lattice}
\phi_{k,m}(\tau, z+\lambda\tau + \mu)
   = \exp\left[-2\pi i m(\lambda^2\tau + 2\lambda z)\right]
   \phi_{k,m}(\tau, z)\, , \quad \lambda,\mu\in \mathbb{Z}\, .
\end{equation}
Furthermore, it has a Fourier expansion
\begin{equation}\label{fj}
    \phi_{k,m}(\tau, z) = \sum_{n\geq 0, \, r\in \mathbb{Z}}c(4n m-r^2)q^{n}
    y^{r}.
\end{equation}
The significance of weak Jacobi forms in this context stems from the
fact that, with the transformation properties (\ref{modular}) and
(\ref{lattice}), the combination $\phi_{k, m} (\rho, \nu) \cdot
\exp(2\pi i m \sigma) $ transforms with weight $k$ under the group
elements $g_1(a, b, c, d)$ and $g_3(\lambda, \mu)$ in
(\ref{egroup}). Then, with some additional ingredients using the
property (\ref{fj}), one can also ensure the required transformation
properties under $g_2$ to obtain a Siegel modular form.

There are two methods for constructing a  Siegel modular form
starting with a weak Jacobi form which we summarize below.
\begin{itemize}
  \item Additive Lift

This  procedure  generalizes the  Maa{\ss}-Saito-Kurokawa lift
explained in detail for example in \cite{Eichler:1985ja}. We refer
to it as the `additive' lift because  it naturally gives the sum
representation of the modular form in terms of its Fourier
expansion. The starting `seed' for the additive lift is in general a
weak Jacobi form $\phi_{k, 1}(\rho, \nu)$ of weight $k$ and index
$1$. Let us denote the operation of additive lift by the symbol
${\cal A}[.]$. If a given weak Jacobi form $\phi_{k, 1}$ results in
a Siegel modular form $\Phi_k$ after the additive lift, then we can
write
\begin{equation}\label{liftrel}
    \Phi_k (\Omega) = {\cal A} [\phi_{k, 1}(\rho, \nu)].
\end{equation}

In the cases of our interest for the pairs $(N, k)$ above, this
procedure was used in \cite{Jatkar:2005bh} to obtain the modular
forms $\Phi_k$ listed there. The seed in these cases can be
expressed in terms of the unique cusp forms $f_k(\rho)$ of
$\Gamma_1(N)$ of weight $(k+2)$,
\begin{equation}\label{cusp}
    f_k(\rho) = \eta^{k+2}(\rho) \eta^{k+2}(N\rho),
\end{equation}
where $\eta(\rho)$ is the Dedekind eta function. The seed for the
additive lift is then given by
\begin{equation}\label{aseed}
    \phi_{k, 1}(\rho, \nu) = f_k(\rho) \, \frac{\theta_1^2(\rho,
    \nu)}{\eta^6(\rho)},
\end{equation}
where $\theta_1(\rho, \nu)$ is the usual Jacobi theta function.

  \item Multiplicative Lift

This procedure is in a sense a logarithmic version of the
Maa{\ss}-Saito-Kurokawa lift. We call it `multiplicative' because it
naturally results in the Borcherds product representation of the
modular form.   The starting `seed' for this lift is a weak Jacobi
form ${\phi}^k_{0,1}$ of weight zero and index one and the
superscript $k$ is added to denote the fact after multiplicative
lift it gives a weight $k$ form $\Phi_k$. Let us denote the
operation of multiplicative lift by the symbol ${\cal M}[.]$. If a
given weak Jacobi form ${\phi}^k_{0,1}$ results in a Siegel modular
form $\Phi_k$ after the multiplicative lift, then we can write
\begin{equation}\label{liftmult}
    \Phi_k (\Omega) =  {\cal M} [\phi^k_{0, 1}(\rho, \nu)].
\end{equation}

\end{itemize}

Given the specific Siegel modular forms $\Phi_k(\Omega)$ obtained by
additively lifting  the seeds $\phi_{k,1}$ in  (\ref{aseed}) for the
pairs $(N, k) = (1, 10), (2, 6), (3, 4), (7, 1)$ as in
\cite{Jatkar:2005bh}, we would like to know if the same Siegel forms
can be obtained as multiplicative lifts of some weak Jacobi forms
$\phi^k_{0,1}$. Such a relation between the additive and the
multiplicative lift is  very interesting mathematically for if it
exists, it gives a Borcherds product representation of a given
modular form.  However, to our knowledge, at present there are no
general theorems relating the two. Fortunately, as we describe next,
in the examples of interest to us, it seems possible to determine
the seed for the multiplicative lift from the seed for the additive
lift quite easily and explicitly. Finding such a multiplicative seed
to start with is a nontrivial step  and is not guaranteed to work in
general. But if one succeeds in finding the multiplicative seed
$\phi^k_{1, 0}$  given a $\Phi_k$ obtained from the  additive seeds
$\phi_{k,1}$ in (\ref{aseed}) then one can write
\begin{equation}\label{final}
   \Phi_k (\Omega) = {\cal A} [f_k(\rho) \, \frac{\theta_1^2(\rho,
    \nu)}{\eta^6(\rho)}] = {\cal M} [\phi^k_{0, 1}(\rho, \nu)].
\end{equation}

\section{Multiplicative Lift \label{Borcherds}}

We now outline  the general procedure for constructing modular forms
$\Phi_k(\Omega)$ as a Borcherds product \cite{Borcherds:1995si} by a
multiplicative lift following closely the treatment in
\cite{Ibukiyama:1991ibu3, Hayashibu:2005sie, AI:2005ai1}

For the special pair $(1, 10)$, which results in the Igusa cusp form
$\Phi_{10}$, the product representation was obtained by Gritsenko
and Nikulin \cite{gritsenko-1996-, Gritsenko:1995ek}. The starting
seed for this lift is a weak Jacobi form ${\phi}^{10}_{0,1}$ of
weight zero and index one
\begin{equation}\label{k3}
    {\phi}^{10}_{0,1} = 8 [ {\theta_2(\rho, \nu)^2 \over \theta_2(\rho)^2} +
    {\theta_3(\rho, \nu)^2 \over \theta_3(\rho)^2} +
    {\theta_4(\rho, \nu)^2 \over \theta_4(\rho)^2}],
\end{equation}
where $\theta_i(\rho, \nu)$ are the usual Jacobi theta functions. We
therefore have in this case the desired result
\begin{equation}\label{rel}
    \Phi_{10} (\Omega) = {\cal A}(\phi_{10,1}) = {\cal
    M}({\phi}^{10}_{0,1}).
\end{equation}
This weak Jacobi form happens to also equal the elliptic genus of
$\bf K3$. As a result, the multiplicative lift is closely related to
the elliptic genus of the symmetric product of $\bf K3$
\cite{Dijkgraaf:1996xw} which counts the bound states of the D1-D5-P
system in five dimensions. This coincidence, which at first sight is
purely accidental, turns out to have a deeper significance based on
the 4d-5d lift \cite{Gaiotto:2005gf}.

We would now like find  a similar product representation for the
remaining pairs of $(N, k)$ using the multiplicative lift so that we
can then try to find a similar physical interpretation using 4d-5d
lift. We first describe the general procedure of the multiplicative
lift for the group $G_0(N)$ and then specialize to the illustrative
case $(2, 6)$ of our interest, to obtain the product representation
of $\Phi_6$ using these methods.

As we have defined in $\S{\ref{Siegel}}$, the  group $G_{0}(N)$
consists of matrices with integer entries of the block-diagonal form
\begin{equation}
\lbrace \left(
\begin{array}{cc}
A & B\\
N C & D\\
\end{array}
\right) \, \in \, Sp(2,\mathbb{Z})\rbrace
\end{equation}
which contains the subgroup $\Gamma_0(N)$. A basic ingredient in the
construction of Siegel modular forms is the Hecke operator $T_{t}$
of $\Gamma_{0}(N)$ where $t$ is an integer. The main property of our
interest is that acting on a weak Jacobi form $\phi_{k,m}$ of weight
$k$ and index $m$, the Hecke operator $T_t$  generates a weak Jacobi
form $\phi_{k,mt} = T_{t}(\phi_{k,m})$ of weight $k$ and index $mt$.
Thus, on a modular form $\phi_{k,1}$, the Hecke operator $T_t$ acts
like   a raising operator that raises the index by $(t-1)$ units.
One subtlety that needs to be taken into account in the case of
$\Gamma_0(N)$ that does not arise for $SL(2, \mathbb{Z})$ is the
fact that $\Gamma_0(N)$ has multiple cusps in its fundamental domain
whereas $SL(2, \mathbb{Z})$ has a unique cusp at $i\infty$. As a
result, the Hecke operators that appear in the construction in this
case are a little more involved as we review in Appendix
\ref{Hecke}.

Let us now explain the basic idea behind the lift. Given a seed weak
Jacobi form $\phi^k_{0,1}(\rho, \nu)$ for the multiplicative lift,
we define
\begin{equation}\label{lphi}
    (L\phi^k_{0,1}) (\rho, \nu, \sigma) =
\sum_{t=1}^{\infty}T_t(\phi^k_{0,1})(\rho,\nu)\exp{(2\pi i \sigma
t)}.
\end{equation}
Now, $T_t(\phi^k_{0, 1})$ is a weak Jacobi form of weight $0$ and
index $t$. It then follows as explained in $\S{\ref{Siegel}}$, with
the transformation properties (\ref{modular}) and (\ref{lattice}),
the combination $T_t( \phi^k_{0, m})(\rho, \nu) \cdot \exp(2\pi i t
\sigma) $ is invariant under the group elements $g_1(a, b, c, d)$
and $g_3(\lambda, \mu)$ in (\ref{egroup}). Thus, each term in the
sum in (\ref{lphi}) and therefor $L\phi$ is also invariant under
these two elements.

If $L\phi$  were invariant also under the exchange of $p$ and $q$
then it would be invariant under the element $g_2$ defined in
(\ref{egroup}) and one would obtain a Siegel modular form of weight
zero. This is almost true. To see this, we note that  $\exp (L
\phi^k_{0,1})$ can be written as an infinite product  using the
explicit representation of Hecke operators given in Appendix
(\ref{Hecke}):
\begin{equation}\label{product}
  \prod_{ l, m, n \in \mathbb{Z} \atop m >0}( 1 -(q^n y^l p^m
  )^{n_s} )^{ h_s n_s^{-1} c_{s, l}(4mn-l^2)},
\end{equation}
where  $ q \equiv \exp(2\pi i \rho), y \equiv \exp(2\pi i \nu), p
\equiv \exp(2\pi i \sigma)$ (\ref{L14}). In the product presentation
(\ref{product}), the coefficients $c_{s, l}(4mn-l^2)$ are manifestly
invariant under the exchange of $m$ and $n$. The product, however,
is not quite symmetric because the range of the products in
(\ref{product}) is not quite symmetric: $m$ is strictly positive
whereas $n$ can be zero. This asymmetry can be remedied by
multiplying the product (\ref{product}) by an appropriate function
as in \cite{gritsenko-1996-2, AI:2005ai1}. The required function can
essentially be determined by inspection to render the final product
symmetric in $p$ and $q$. Following this procedure one then obtains
a Siegel modular form as the multiplicative lift of the weak Jacobi
form $\phi^k_{0,1}(\rho, \nu)$,
\begin{equation}\label{final2}
  \Phi_k(\Omega) = {\cal M} [\phi^k_{0,1}] = q^a y^b p^c \prod_{(n, l, m) >0}( 1 -(q^n y^l p^m
  )^{n_s} )^{ h_s n_s^{-1} c_{s, l}(4mn-l^2)},
\end{equation}
for some integer $b$ and positive integers $a, c$. Here the notation
$(n, l, m) >0$ means that if (i) $m >0, n, l \in \mathbb{Z}$, or
(ii) $m=0, n >0, l\in \mathbb{Z}$, or (iii) $m= n= 0, l < 0$.

It is useful to write  the final answer for $\Phi_k(\Omega)$ as
follows
\begin{eqnarray}\label{final2hodge}
  \Phi_k(\Omega) &=& p^c H(\rho, \nu) \exp[L\phi^k_{0,1}(\rho, \nu,
  \sigma)],
\end{eqnarray}
\begin{eqnarray}\label{hodge2}
  H(\rho, \nu)&=& q^a y^b \prod_s \prod_{l, n \geq 1} ( 1- (q^n y^l)^{n_s})
  ( 1- (q^n y^l)^{n_s})^{n_s^{-1} h_s c_{s, l}(-l^2)}\\
  &\times& \prod_{n=1}^\infty ( 1- q^{n n_s})^{n_s^{-1}h_s
  c_{s,l}(0)} \prod_{l < 0}^\infty ( 1- y^{l n_s})^{n_s^{-1}h_s
  c_{s,l}(-l^2)},
\end{eqnarray}
in terms of the separate ingredients that go into the construction.
This rewriting is more suggestive for the physical interpretation,
as we discuss in the next section. Following Gritsenko
\cite{gritsenko-1999-}, we refer to the function $H(\rho, \nu)$ as
the `Hodge Anomaly'. The construction thus far is general and
applies to the construction of modular forms of weight $k$ which may
or may not be obtainable by an additive lift. In many cases however,
as in the cases of our interest, it might be be possible to obtain
the same modular form by using the two different lifts. To see the
relation between the two lifts in such a situation and to illustrate
the significance of the Hodge anomaly for our purpose, we next
specialize to the case $(2, 6)$. We show how to determine the
multiplicative seed and the Borcherds product given the specific
$\Phi_6$ obtained from the additive lift.

\section{Multiplicative Lift for $\Phi_{6}$ \label{Example}}

We want to determine the seed $\phi^6_{0, 1}$ whose multiplicative
lift  equals $\Phi_6$ constructed from the additive lift of
(\ref{aseed}). From the $p$ expansion of the additive representation
of $\Phi_6$ we conclude that the integer $c$ in (\ref{final2}) and
(\ref{final2hodge}) equals one. Then we see from (\ref{final2hodge})
that if $\Phi_6$ is to be a weight six Siegel modular form, $H(\rho,
\nu)$ must be a weak Jacobi form of weight six and index one. Such a
weak Jacobi form is in fact unique and hence must equal the seed
$\phi_{6, 1}$ that we used for the additive lift. In summary, the
Hodge anomaly is given by
\begin{eqnarray}
  H(\rho, \nu)  &=& \phi_{6,1}(\rho, \nu)=\eta^{2}
(\rho) \eta^{8}(2\rho)\theta_{1}^{2}(\rho, \nu) \\
   &=& q y (1 - y^{-1})^2\prod_{n=1}^\infty(1- q^{2n})^8 (1-q^n)^4 (1-q^n y)^2
   (1-q^ny^{-1})^2. \label{hodgeprod}
\end{eqnarray}
Comparing this product representation with (\ref{hodge2}), we
determine that
\begin{equation}\label{coeff}
    c_1(0) = 4, \quad c_1(-1) =2; \quad c_2(0) =8, \quad c_2(-1) =0;
\end{equation}
and moreover $c_1(n)=c_2(n) =0, \quad \forall\, n <-1$. This
information about the leading coefficients $c_s(n)$ obtained  from
the Hodge anomaly is sufficient to determine completely the
multiplicative seed ${\phi}^6_{0,1}$. Let us assume the seed to be a
weak Jacobi form\footnote{Strictly, it is enough that it is a `very
weak' Jacobi form as defined in \cite{AI:2005ai1} but from the
physical interpretation  that we give in the next section, we expect
and hence assume it to be a weak Jacobi form to find a consistent
solution.}. Now, proposition (6.1) in \cite{AI:2005ai1} states that
the space of weak Jacobi forms of even weight  is generated as
linear combinations of two weak forms $\phi_{-2,1}$ and $\phi_{0,1}$
which in turn are given in terms of elementary theta functions by
\begin{eqnarray}\label{ring}
   \phi_{-2,1}(\rho, \nu) = \frac{\theta_1^2(\rho,
    \nu)}{\eta^6(\rho)}  \\
  \phi_{0,1}(\rho, \nu)= 4 [ {\theta_2(\rho, \nu)^2 \over \theta_2(\rho)^2} +
    {\theta_3(\rho, \nu)^2 \over \theta_3(\rho)^2} +
    {\theta_4(\rho, \nu)^2 \over \theta_4(\rho)^2}]
\end{eqnarray}
The coefficients for this linear combination can take values in the
ring $A(\Gamma(N))$ of holomorphic modular forms of $\Gamma(N)$.
Basically, the coefficients have to be chosen so as to get the
correct weight. For our case, with $N=2$, the relevant holomorphic
modular form, is the one of weight two
\begin{equation}\label{alpha}
\alpha(\rho)=\theta_{3}^4 (2\rho)+\theta_{2}^4(2\rho).
\end{equation}
By virtue of the above-mentioned proposition, and using the
definitions in (\ref{ring}) and (\ref{alpha}), we can then write our
desired seed as the linear combination
\begin{equation}\label{linear}
\phi^6_{0, 1}(\rho, \nu) = A  \alpha(\rho) {\phi}_{-2,1}(\rho, \nu)
+ B {\phi}_{0, 1}(\rho, \nu),
\end{equation}
where $A$ and $B$ are constants. To determine the constants we
investigate the behavior near the cusps. For $\Gamma_0(2)$, there
are only two cusps, one at $i\infty$ and the other at $0$ in the
fundamental domain which we label by $s=1, 2$ respectively. Then the
various relevant quantities required in the final expression
(\ref{final2}) are given in our case by
\begin{eqnarray}
  g_1 = \left(
          \begin{array}{cc}
            1 & 0 \\
            0 & 1 \\
          \end{array}
        \right)
   &\quad & h_1 =1, \quad  z_1 =0,\quad n_1 = 1\\
  g_2 = \left(
          \begin{array}{cc}
            0& -1 \\
            1& 0 \\
          \end{array}
        \right)
   &\quad & h_1 = 2, \quad  z_2 =1,\quad n_2 = 2.
\end{eqnarray}
The $q$ expansion for $\phi_{-2,1}$ and $\phi_{0,1}$ at the cusp
$q=0$ is given by
\begin{equation}\label{phiexp1}
\phi_{-2,1}= (-2+y+y^{-1})+q(-12+8y+8y^{-1}-2y^{2}-2y^{-2})+.....
\end{equation}
\begin{equation}\label{phiexp2}
\phi_{0,1}=(10+y+y^{-1})+\,....
\end{equation}
The Fourier expansion of $\alpha(\rho)$ at the cusps $i\infty$ and
$0$ is given by,
\begin{eqnarray}\label{alphaexp1}
  \alpha(\rho) &=& 1 + 24 q + 24 q^2 + \ldots
\end{eqnarray}
near infinity and by
\begin{eqnarray}\label{alphaexp2}
  \rho^{-2} \alpha({-1}/{\rho}) &=& -\frac{1}{2}\alpha(\frac{\rho}{2})\\
                              &=&  -\frac{1}{2} + \ldots
\end{eqnarray}
near zero. Demanding that the leading terms in the Fourier expansion
of the linear combination (\ref{linear}) match with those given by
(\ref{coeff}) determines the coefficients $A= 4/3$ and $B= 2/3$ in
(\ref{linear}). The constraints are actually over-determined so the
fact that a solution  exists at all gives a check of the procedure.
Our final answer for the multiplicative lift is then
\begin{equation}\label{seed}
    \phi^6_{0, 1}(\rho, \nu) = {4\over 3} \alpha(\rho)
    {\phi}_{-2,1}(\rho, \nu) + {2\over 3}
    {\phi}_{0,1}(\rho, \nu).
\end{equation}
With this determination we can simply apply the formalism in the
previous section to determine
\begin{equation}\label{mathfinal}
    \Phi_6(\Omega) = {\cal M} [{4\over 3} \alpha(\rho)
    {\phi}_{-2,1}(\rho, \nu) + {2\over 3}
    {\phi}_{0, 1}(\rho, \nu)]
\end{equation}
by using the formula (\ref{final2}).

\section{Physical Interpretation of the Multiplicative Lift
\label{Interpret}}

Both $\exp (-L\phi)$ and the inverse of the  Hodge anomaly
$H^{-1}(\rho, \nu)$ that appear in the multiplicative lift in
(\ref{Borcherds}) have a natural physical interpretation  using the
4d-5d lift, which we discuss in this section and also in terms of
M-theory lift of string webs which we discuss in the next section.

Let us recall the basic idea behind  the 4d-5d lift
\cite{Gaiotto:2005gf}. Consider Type-IIA compactified on a
five-dimensional space $\bf M_5$ to five dimensions. Given a BPS
black hole in Type-IIA string theory in five dimensions, we can
obtain a black hole in four dimensions as follows. A
five-dimensional black hole situated in an asymptotically flat space
$\mathbb{R}^4$ can be embedded into an asymptotically Taub-NUT
geometry of unit charge. Intuitively, this is possible because near
the origin, the Taub-NUT geometry reduces to $\mathbb{R}^4$, so when
the Taub-NUT radius is much larger than the black hole radius, the
black hole does not see the difference between $\mathbb{R}^4$ and
Taub-NUT. Asymptotically, however, the Taub-NUT geometry is
$\mathbb{R}^3 \times \bf{S_{tn}^1}$. We can dimensionally reduce on
the Taub-NUT circle to obtain a four-dimensional compactification.
Now, Type-IIA is dual to M-theory compactified on the M-theory
circle $\textbf{S}^1_m$ so we can regard four-dimensional theory as
an M-theory compactification on $\bf M_5 \times S_{tn}^1 \times
S^1_m$. Now flipping the two circles, we can choose to regard the
Taub-NUT circle $\bf S_{tn}^1$  as the new M-theory circle. This in
turn is dual to a Type-IIA theory but in a different duality frame
than the original one. In this duality frame, the Taub-NUT space is
just the Kaluza-Klein 6-brane of M-theory dual to the D6-brane. Thus
the Taub-NUT charge of the original Type-IIA frame is interpreted in
as the D6 brane charge in the new Type-IIA frame and we obtain a BPS
state in four dimensions with a D6-brane charge. Since we can go
between the two descriptions by smoothly varying various moduli such
as the Taub-NUT radius and choosing appropriate duality frames, the
spectrum of BPS states is not expected to change. In this way, we
relate the spectrum of four-dimensional BPS states with D6-brane
charge to five-dimensional BPS states in Type-IIA.

With this physical picture in mind, we now interpret the term
$\exp(-L\phi)$ in (\ref{final2hodge}) as counting the degeneracies
of the five dimensional BPS states that correspond to the
four-dimensional BPS states after the 4d-5d lift. For example, in
the familiar case $(1, 10)$ of toroidally compactified heterotic
string, the dual Type-II theory is compactified on $\bf K3 \times
\tilde{S}^1 \times S^1$. In the notation of the previous paragraph,
we then have $\bf M_5 = K3 \times \tilde{S}^1$. The five-dimensional
BPS state is described by the  D1-D5-P system. Its degeneracies are
counted by the elliptic genus of the symmetric product of $\bf K_3$.
In this case, indeed $\exp(-L\phi)$ above gives nothing but the
symmetric product elliptic genus evaluated in
\cite{Dijkgraaf:1996xw}.

In our case $(2, 6)$, D-brane configuration in five dimensions
corresponding to our dyonic state in four dimensions is obtained
simply by implementing the CHL orbifolding action in the open string
sector on the D1-D5-P system in five dimensions. The term
$\exp(-L\phi)$ in (\ref{final2hodge}) then has a natural
interpretation as a symmetric product elliptic genus. Because of the
shift in the orbifolding action, the resulting orbifold is a little
unusual and the elliptic genus is weak Jacobi form not of $SL(2,
\mathbb{Z})$ but of $\Gamma_0(2)$. The details of the orbifold
interpretation will be presented in \cite{Dabholkar:2006}.

The Hodge anomaly plays a special role in the 4d-5d lift. It is
naturally interpreted as the contribution of the bound states of
momentum and the single Taub-NUT 5-brane in the Type-IIB
description. A KK5-brane of IIB wrapping $\bf K_3 \times S^1$
carrying momentum along the $\bf S^1$  is T-dual to an NS5-brane of
IIA wrapping $\bf K_3 \times S^1$ carrying momentum which in turn is
dual to the heterotic fundamental string wrapping the circle with
momentum.\footnote{In
\cite{Shih:2005qf}, the Hodge anomaly for the $(1, 10)$ example is
interpreted as a single 5-brane contribution. This, however, is not
dual to the heterotic F1-P system and would not give the desired
form of the Hodge anomaly. For the purposes of 4d-5d lift, it is
essential to introduce Taub-NUT geometry which appears like
KK5-brane in IIB. In the 5d elliptic genus the bound states of this
KK5-brane and momentum are not accounted for. Therefore, the Hodge
anomaly is naturally identified as this additional contribution that
must be taken into account.}. These can be counted in perturbation theory
\cite{Dabholkar:1989jt,Dabholkar:1990yf,Dabholkar:2005by,Dabholkar:2005dt}
in both cases $(1, 10)$ and $(2, 6)$. The $y(1-y^{-1})$ term in the Hodge anomaly in (\ref{hodgeprod}) is more subtle and would require a more detailed analysis.

\section{M-theory lift of String Webs  \label{Web}}

The appearance in the dyon counting formulae of objects related to a
genus two Riemann surface such as the period matrix and the $G_0(N)$
subgroups of $Sp(2, \mathbb{Z})$ is quite surprising and demands a
deeper physical explanation. We now offer such an explanation
combining earlier work of \cite{kawai-1997-} and
\cite{Gaiotto:2005gf} in the toroidal $(1, 10)$  case and
generalizing it to CHL orbifolds.

To start with, let us reinterpret the Hodge anomaly following Kawai
\cite{kawai-1997-}. It can be written as
\begin{equation}\label{rehodge}
    H(\rho, \nu) = \eta^{8}
(\rho) \eta^{8}(2\rho)\frac{\theta_{1}^{2}(\rho, \nu)}{\eta^6(\rho)}
= Z(\rho) K^2(\rho, \nu),
\end{equation}
where $Z(\rho)\equiv \eta^{8} (\rho) \eta^{8}(2\rho)$ is the
one-loop partition function of the left-moving chiral 24-dimensional
bosonic string with the $\mathbb{Z}_2$ twist $\alpha$ of the CHL
orbifold action, and $K(\rho, \tau)$ is the prime form on the torus.
Let us also expand
\begin{equation}\label{ell}
    \exp(-L\phi^6_{0,1}(\rho, \nu)) = \sum_{N=0}^\infty p^N \chi_N
\end{equation}
We can then write from (\ref{final2hodge}),
\begin{eqnarray}\label{oneover}
 \frac{1}{\Phi_6(\Omega)} &= & \frac{1}{p} \frac{1}{H(\rho, \nu)}
    \exp(-L\phi^6_{0,1}(\rho, \nu)) \\
   &=& \sum_{N=0}^\infty p^{N-1} \frac{1}{K(\rho, \nu)^2} \chi_N\\
    &\sim&  \frac{1}{p} \frac{1}{K(\rho, \nu)^2} \frac{1}{Z(\rho)} +
    \ldots \label{tachpole}
\end{eqnarray}
\begin{figure}
\centering
\includegraphics[width=1.2in,angle=270]{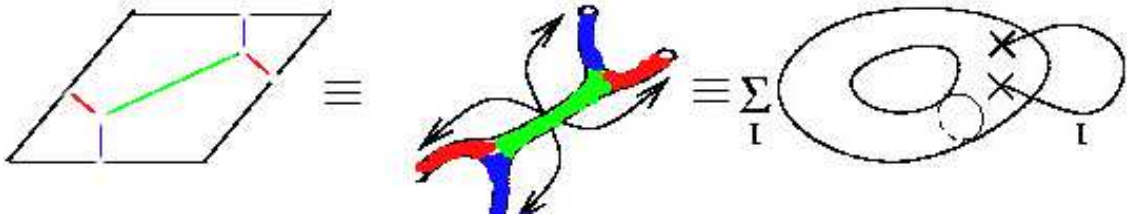}
\caption{A dyon can be represented as a string web on a torus which
in M-theory looks like a genus two Riemann surface.  Factorization
of the product representation of $1/\Phi_k(\Omega)$ reveals this
Riemann surface.} \label{twoloop}
\end{figure}
In (\ref{tachpole}) above, we can identify $K^{-2}(\rho, \nu)$ as
the on-shell (chiral) tachyon propagator, and $Z(\rho)$ as the
one-loop left-moving partition function. If we denote by $X$ the
bosonic spacetime coordinate, then we have
\begin{equation}\label{tach}
    <e^{i k \cdot X}(\nu) e^{-i k \cdot X}(0)> = K^{-2} (\rho, \nu),
\end{equation}
where $k$ is the momentum  of an on-shell tachyon and the correlator
is evaluated on a genus one Riemann surface with complex structure
$\rho$. This is  exactly  the first term in the factorized expansion
in Fig.(\ref{twoloop}). The subleading terms at higher  $N$ denoted
by $\ldots$ in (\ref{tachpole}) come from contributions of string
states at higher mass-level $N-1$. Summing over all states gives the
genus two partition function.

This reinterpretation of $1/\Phi_6$ as the two-loop partition
function of the chiral bosonic string explains at a mathematical
level the appearance of genus two Riemann surface generalizing the
results of Kawai to the $(2, 6)$ case.  Note that the partition
function $Z(\rho)$ will be different in the two cases. In the (1,
10) case it equals $\eta^{-24}(\rho)$ and in the $(2, 6)$ case it
equals $\eta^{-8}(\rho) \eta^{-8}(2 \rho)$. This precisely captures
the effect  of CHL orbifolding on the chiral left moving bosons of
the heterotic string. To describe the $N=2$ orbifold action let us
consider the $E_8 \times E_8$ heterotic string. The orbifold twist
$\alpha$ then simply flips the two $E_8$ factors. We can compute the
partition function with a twist in the time direction $
\textrm{Tr}(\alpha q^H)$ where $H$ is the left-moving bosonic
Hamiltonian. Then, the eight light-cone bosons will contribute
$\eta^{-8}(\rho)$ as usual to the trace, but the sixteen bosons of
the internal $E_8 \times E_8$ torus will contribute
$\eta^{-8}(2\rho)$ instead of $\eta^{-16}(2\rho)$. The power changes
from $-16$ to $-8$ because eight bosons of the first $E_8$ factor
are mapped by $\alpha$ to the eight bosons of the second $E_8$. Thus
only those states that have equal number of oscillators from the two
$E_8$ factors contribute to the trace, thereby reducing effectively
the number of oscillators to $8$. The argument on the other hand is
doubled  to $2 \rho$ because when equal number of oscillators from
the two factors are present, the worldsheet energy is effectively
doubled. The tachyon propagator in the two cases is unchanged
because only light-cone bosons appear on shell which are not
affected by the orbifolding.

This mathematical rewriting  does not explain at a fundamental level
why the chiral bosonic string has anything to do with dyon counting.
This connection can be completed using with the heuristic picture
suggested in \cite{Gaiotto:2005hc}.

Under string-string duality, the $SL(2, \mathbb{Z})$  S-duality
group of the heterotic string gets mapped to the geometric $SL(2,
\mathbb{Z})$ of the Type-IIB string
\cite{Hull:1995nu,Witten:1995ex}. Thus, electric states correspond
to branes wrapping the $a$ cycle of the torus and magnetic states
correspond to branes wrapping the $b$ cycle of the torus. A general
dyon with electric and magnetic charges  $(n_e, n_m)$ of a given
$U(1)$ symmetry is then represented as a brane wrapping $(n_e, n_m)$
cycle of the torus. If a state is charged under more than one $U(1)$
fields then one gets instead a $(p, q)$ string web with different
winding numbers along the $a$ and the $b$ cycles. The angles and
lengths of the web are fixed by energetic considerations for a given
charge assignment  \cite{Sen:1997xi, Aharony:1997bh}. For our
purpose, we can consider D5 and NS5 branes wrapping the $\bf K3$
resulting in two different kinds of $(1, 0)$ and $(0, 1)$ strings. A
dyon in a particular duality frame then looks like the string web
made of these strings as in the first diagram Fig.(\ref{twoloop}).
In the M-theory lift of this diagram, both D5 and NS5 branes
correspond to M5 branes so the string in the web arises from
M-theory brane wrapping $\bf K3$. To count states, we require a
partition function with Euclidean time. Adding the circle direction
of time we can fatten the string web diagram which looks like a
particle Feynman diagram into a genus-two Riemann surface
representing a closed-string Feynman diagram as in the second
diagram in Fig.(\ref{twoloop}). Now, $\bf K3$-wrapped M5 brane is
nothing but the heterotic string. Furthermore, since we are counting
BPS states by an elliptic genus, the right-movers are in the ground
state and we are left with the two-loop partition function of the
bosonic string. This partition function is what we have constructed
by in the third diagram in  Fig.(\ref{twoloop}) as explained above
using factorization.

\section{Conclusions}

The exact spectrum of dyons in four dimensions and of spinning black
holes in five dimensions in CHL compactifications can be determined
using a Borcherds product representation of level $N$ Siegel modular
forms of $Sp( 2, \mathbb{Z})$. Various elements in the Borcherds
product have a natural interpretation from the perspective of 4d-5d
lift. The Hodge anomaly is identified the contribution of bound
states of Type-IIB KK5-brane with momentum. The remaining piece is
interpreted as arising from the symmetric product of  the orbifolded
D1-D5-P system. The appearance of an underlying chiral bosonic
string on a genus two Riemann surface in this construction has a
natural interpretation  as the Euclidean worldsheet of the  $\bf K3$
wrapped M5 brane on a string web in orbifolded theory. By
factorization, this connection with the Siegel modular form can be
made precise.

We have seen that a very rich and interesting mathematical structure
underlies the counting of BPS dyons and black holes. Given the
relation of Siegel modular forms to Generalized Kac-Moody algebras
\cite{Borcherds:1995si, gritsenko-1996-1, gritsenko-1996-2,
Harvey:1995fq}, their appearance in the counting is perhaps
indicative of a larger underlying symmetry of string theory. If so,
investigating this structure further might prove to be a fruitful
avenue towards uncovering the full structure of M-theory.

\textsl{Note:} During the course of writing this paper, a related
paper appeared \cite{David:2006ji} with some overlap with our work
where the product representation is derived using yet another lift
called the `Theta Lift'.

\subsection*{Acknowledgements}

We would like to thank Eknath Ghate, Dileep Jatkar, Arvind Nair,
Dipendra Prasad, Ashoke Sen, and T. Venkataramana for useful
discussions. We are especially grateful to Davide Gaiotto for
explaining his paper and to  Rainer Schulze-Pillot for explaining
some elements of the Borcherds lift.

\appendix

\section{Hecke Operators and the Multiplicative Lift \label{Hecke}}
In this section we summarise the construction of Hecke operators and the multiplicative lift, following \cite{AI:2005ai1}. 
Let us define $\Delta_N(t)$ as
\begin{equation}\label{delta}
   \Delta_N(t) = \{ g =  \left(
                           \begin{array}{cc}
                             a & b \\
                             cN  & d \\
                           \end{array}
                         \right); \quad a, b, c, d \in \mathbb{Z},
                         \quad \det(g) =t
   \}.
\end{equation}
The action of the Hecke operator $T_t$ on a weak Jacobi form
$\phi_{k,m}$ is then  given by
\begin{equation}
 T_{t}(\phi_{k,m})(\tau, z)=t^{k-1}\sum_{
\left(
\begin{array}{cc}
a & b\\
c & d \\
\end{array}
\right) \,\in\,
\Gamma_{0}(N)\backslash\Delta_{N}(t)}(c\tau+d)^{-k}\exp{(-\frac{2
\pi i mc z^2}{c\tau+d})}\phi_{k,m}(\frac{a\tau+b}{d},az).
\end{equation}
To compute everything concretely, we need to define representatives
of $\Gamma_{0}(N)\backslash\Delta_{N}(t)$. Choose
the complete set of cusps $\{s\}$ of $\Gamma_0(N)$ represented by
the set of representative matrices $\{ g_s\}$. Let
 \begin{equation}
g_{s}\,\epsilon\,SL(2,\mathbb{Z})= \left(
\begin{array}{cc}
x_{s} & y_{s}\\
z_{s} & w_{s}\\
\end{array}
\right)
\end{equation}
Define a natural number $h_{s}$ by
\begin{center}
$ g_{s}^{-1}\Gamma_{0}(N)g_{s}\cap P(\mathbb{Z})=\lbrace\pm \left(
\begin{array}{cc}
1 & h_{s}n\\
 0 & 1\\
\end{array}
\right) ;n\,\in\,\mathbb{Z}\rbrace
$\\
\end{center}
where $P(\mathbb{Z})$ is the set of all upper-triangular matrices
over integers with unit determinant. We can then write
 \begin{equation}
\Gamma_{0}(N)\backslash \Delta_{N}(t)=\cup_{s}\lbrace{g_{s} \left(
\begin{array}{cc}
a & b\\ 0 & d\\
\end{array}
\right);\, a,b,d\,\in\, \mathbb{Z},ad=t, \, az_{s}=0 \, {mod} \,
N,\, b=0,...,h_{s}d-1}\rbrace.
\end{equation}
For each cusp we  define $n_s = \frac{N}{\textrm {g.c.d}(z_s , N)}$.
We define
\begin{equation}\label{phis}
    \phi_s(\tau, z) = \phi_( \frac{x_s \tau + y_s}{z_s \tau +
    \omega_s}, \frac{z}{z_s \tau +
    \omega_s}),
\end{equation}
with Fourier expansion
\begin{equation}\label{phifour}
     \phi_s(\tau, z) = \sum_{n, l} c_s(n, l) \exp(2 \pi i ( n\tau +
     l z)).
\end{equation}
As usual, one can show that $c_s(n, l)$ depends only on $4n - l^2$
and $ l\, mod \, 2$ so we write $c_s(n, l) = c_{s, l}(4n-l^2)$
following the notation in \cite{AI:2005ai1}. In general $n \in
h_s^{-1} \mathbb{Z}$ need not be an integer. If $4$ does not divide
$h_s$, which is true  for all cases of our interest, then $l\, mod\,
2$ is determined only by $4n -l^2$ and in that case we can write
simply $c_s(4n -l^2) = c_{s, l} (4n -l^2)$.

For $\Gamma_0(N)$ with $N$ prime, there are only two cusps, one at
$i\infty$ and the other at $0$ in the fundamental domain. Hence the
index $s$ runs over 1 and 2. For this case, various objects with the
subscript $s$ defined in the formula for the lift above take the
following values:
\begin{eqnarray}
  g_1 = \left(
          \begin{array}{cc}
            1 & 0 \\
            0 & 1 \\
          \end{array}
        \right)
   &\quad & h_1 =1, \quad  z_1 =0,\quad n_1 = 1\\
  g_2 = \left(
          \begin{array}{cc}
            0& -1 \\
            1& 0 \\
          \end{array}
        \right)
   &\quad & h_2 =N, \quad  z_2 =1,\quad n_2 = N
\end{eqnarray}
In this case we can then write
\begin{eqnarray}
&& \Gamma_{0}(N)\backslash \Delta_{N}(t)= \lbrace \left(
\begin{array}{cc}
a & b\\
0 & d\\
 \end{array}
\right) \, \in \, GL( 2, \mathbb{Z});ad=t,b=0,...,d-1 \rbrace \\
&\cup& \lbrace  g_{2} \left(
\begin{array}{cc}
a & b\\
0 & d\\
\end{array}
\right) \,\in \,GL( 2, \mathbb{Z});ad=t,a\equiv 0\, mod \, N,
b=0,....,Nd-1 \rbrace.
\end{eqnarray}
Given a weak Jacobi form $\phi$ of weight $0$ and index $1$, we can
define
\begin{equation}\label{L}
    L\phi (\rho, \nu, \sigma) =
\sum_{t=1}^{\infty}T_t(\phi)(\rho,\nu)\exp{(2\pi i \sigma t)}.
\end{equation}
Using the explicit representation of the Hecke operators, one can
then show \cite{AI:2005ai1}
\begin{eqnarray}
 L\phi &=& \sum_s \sum_{t=1}^{\infty} \sum _{ad =t \atop az_s = 0\, {mod}\, N}\sum_{b=0}^{h_s d -1}
           \phi_s(\frac{a \rho + b}{d}, a\nu) \exp(2\pi i t \sigma) \\\label{L11}
   &=& \sum_s \sum_{t=1}^{\infty} \sum _{ad =t \atop a \in n_s \mathbb{Z}}
   (ad )^{-1} d h_s \sum_{n, l \in  \mathbb{Z}} c_{s, l} (4 nd -l^2)
   \exp(2\pi i ( an \rho + a l\nu + t \sigma))\\\label{L12}
   &=& \sum_s h_s \sum_{a=1}^\infty \frac{1}{an_s} \sum_{m=1}^{\infty} \sum_{n, l \in
   \mathbb{Z}} c_{s, l} (4 nd -l^2)
   \exp(2\pi i ( an \rho + a l\nu + m \sigma))\\\label{L13}
   &=& \sum_{s}\frac{h_{s}}{n_{s}}\log\left(\prod_{l,m,n\, \in
\, \mathbb{Z}\atop
m\geq1}(1-e^{n_{s}(n\rho+l\nu+m\sigma)})^{c_{s,l}(4mn-l^{2})}\right).\label{L14}
\end{eqnarray}

\section{Consistency Check \label{check}}

As a consistency check we compare the coefficients of the leading
powers of $p, q, y$ in the multiplicative lift with the Fourier
expansion of $\Phi_{6}$ obtained using the additive lift in
\cite{Jatkar:2005bh}. The leading terms, corresponding to a single
power of $p$, in the expansion are
\begin {equation}
-pqy\prod_n(1-q^{n})^{c_{1}(0)}(1-q^{n}y)^{c_{1}(-1)}(1-q^n
y^{-1})^{c_{1}(-1)}(1-q^{2n})^{c_{2}(0)}.
\end{equation}
Substituting the values of the $c_1$ and $c_2$  coefficients and
collecting terms with the same powers in q and y together, we obtain
\begin{equation}\label{expfi}
\Phi_{6}(\Omega) =[(2-y -\frac{1}{y})q + (-4+\frac{2}{y^2})q^{2} +
(-16 -\frac{1}{y^{3}})-\frac{4}{y^2}+\frac{13}{y} + 13y
-4y^{2}-y^{3})q^{3}]p + \ldots
\end{equation}
To compare, we now read off the coefficients from its sum
representation  derived in \cite{Jatkar:2005bh} by the additive
lift. The seed for the additive lift is
\begin{equation}\label{adfo}
\phi_{6,1}=\eta^{2}(\tau) \eta^{8}(2\tau)\theta_{1}^{2} = \sum_{l ,
n\geq 0} C(4n -l^2) q^n y^l
\end{equation}
The lift is then given by
\begin{equation}
\Phi_{6}(\Omega) = \sum_{m \geq 1} T_{m}[\phi_{6,1}(\rho, \nu)]
p^{m},
\end{equation}
with the Fourier expansion
\begin{equation}\label{four}
  \Phi_{6}(\Omega)  = \sum_{m >0, \atop {n \geq 0, r \in \mathbb{Z}}}a(n,m,r) q^n p^m
  y^r.
\end{equation}
Given the action of the Hecke operators, $a(n, m, r)$ can be read
off from this expansion knowing $C(N)$  as in (\ref{adfo}). These
are in precise agreement with the same coefficients in the expansion
of the product representation given above in (\ref{expfi}).

\bibliographystyle{utphys}
\bibliography{ref2}

\end{document}